\documentstyle[12pt,a4,epsf]{article}
\def \half {{\textstyle1\over2}}
\def \Trans{{\scriptscriptstyle T}}
\def \kT   {k_{\scriptscriptstyle T}}
\def \pT   {p_{\scriptscriptstyle T}}
\def \QT   {q_{\scriptscriptstyle T}}
\def \fcol {f_{\mathrm{col}}}
\def \gev  {\mbox{$\,$GeV}}
\def \nb   {\mbox{$\,$nb}}
\def \topline#1{\raisebox{48pt}[0pt][0pt]{\makebox[\textwidth]{#1}}\relax}
\let \captionsize=\small
\begin{document}

\title {
        \topline{\hfill MITH 94/13}\\
	The Pomeron in QCD\\
       {\normalsize (submitted to {\em Phys. Lett.} {\bf B}.)}
}

\author{
        Giuliano PREPARATA\\
       {\small\it Dip.~di Fisica, Univ.~degli Studi,
                  via Celoria 16, I-20133 Milano, Italy}\\
       {\small    and}\\
       {\small\it INFN - Sezione di Milano
                  via Celoria 16, I-20133 Milan, Italy}
\and
	Philip G.~RATCLIFFE\\
       {\small\it Dip.~di Fisica, Univ.~degli Studi,
                  via Celoria 16, I-20133 Milano, Italy}\\
}

\date{September 1994}

\maketitle

\begin{abstract}
In the framework of Anisotropic Chromodynamics, a non-perturbative realization
of QCD, we develop the Low-Nussinov picture of the Pomeron. In this approach
all the usual problems of low $\pT$ perturbative calculations (infrared
divergence) are naturally absent. Thus, we are able to perform an {\em ab
initio\/} calculation of the hadron-hadron total cross section. The result is
a cross section of the same magnitude as indicated experimentally and
approximately energy-independent (with a $\log^2s$ growth). We further discuss
the $\pT$ dependence of the hadron-hadron elastic-scattering cross section,
which displays all the experimentally observed features. \end{abstract}

\section {Introduction}

Ever since it was realized that hadrons are extended objects and that hadronic
interactions at high energies have a finite (transverse) size, the natural
description of high-energy scattering at small angles has always been in terms
of diffraction. While the intuitive notion of diffraction is very easy to
understand, the actual dynamical mechanisms that render a hadron a kind of
``grey disk'' at high energies have in the last thirty years been the subject
of the most diverse speculation.

Under the hegemony of ``Reggeology'' during the sixties, when dynamics and
(angular momentum) analyticity were thought to be tied together in a sort of
one-to-one correspondence, it was only natural to look for the
description/explanation of the observed (approximate) constancy of hadronic
cross sections in the high-energy asymptotic limit in terms of a Regge
trajectory with intercept $\alpha_{I\!\!P}(0)=1$, the Pomeron \cite{Pom}. With
the subsequent advent of higher energy accelerators (ISR, Fermilab,
$\mathrm{Sp\bar{p}S}$ at CERN, the Tevatron collider) it became ever clearer
that the Pomeron singularity in the complex angular-momentum plane is quite
different from that of the subleading Regge trajectories, whose connection with
the low-lying meson spectrum completely corroborates the theoretical
foundations of the Regge-pole approach to hadrodynamics. In particular, no
meson has ever been found to be associated with the Pomeron trajectory. It thus
became clear that the simple high-energy dynamics of Regge poles, relating a
``tower'' of $t$-channel exchanges to the characteristic power behaviour in $s$
(the CM energy squared) of the high-energy scattering amplitudes, is totally
inadequate. But, if not a Regge pole, what then is the Pomeron?

\begin{figure}[hbt]
\centering
\epsfysize=5cm
\mbox{\epsfbox{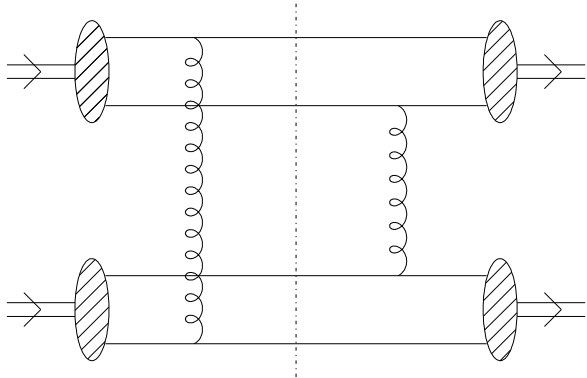}}
\caption{\captionsize The Low-Nussinov picture of the Pomeron as a two-gluon
exchange process (all permutations of the quark-gluon vertices should be
summed). }
\label{fig:2gluon}
\end{figure}

The answer that, in our opinion, came closer to the truth was provided in the
middle of the seventies by the very interesting proposal of F.~Low and
S.~Nussinov \cite{LowNus}. According to these authors, if one considers the QCD
Feynman diagrams of fig.~\ref{fig:2gluon}, where the interaction is the simple
perturbative one-gluon exchange between quarks (obviously modified by
appropriate non-perturbative form factors) then the related hadronic amplitudes
would exhibit the observed high-energy behaviour. In particular, the structure
of the angular-momentum plane singularity associated with the ``two-gluon
exchange'' diagrams of fig.~\ref{fig:2gluon} is certainly not that of a pole --
the Pomeron trajectory -- but rather that of a (possibly fixed) cut.

We should like to remark that the definite achievement of the Low-Nussinov
proposal lies in its focussing attention on a simple dynamics of the hadronic
fundamental degrees of freedom -- quarks and gluons -- showing that very
interesting and sensible results can be obtained even in the regime where
perturbative QCD (PQCD) is deemed to be completely inapplicable, owing to the
outstanding problem of confinement. On the other hand, taken at face value, the
diagrams of fig.~\ref{fig:2gluon} must be considered absolutely {\em
unrealistic}, for their imaginary parts, which build up most of the high-energy
amplitudes, arise from intermediate states in which there freely propagate
coloured degrees of freedom -- quarks and gluons -- contrary to observations
that attribute those imaginary parts (through the optical theorem) to rather
complicated multihadron intermediate states.

In this letter we wish to show that in the Anisotropic Chromodynamics (ACD)
realization of non-perturbative QCD one can develop the seminal idea of Low and
Nussinov in a completely consistent way, without the grave difficulties we have
just pointed out. Without entering into a detailed description of ACD and how
it describes non-perturbatively QCD, space does not permit, we wish to remind
the reader that the theoretical framework we shall adopt in this paper has been
derived from a detailed analysis of the structure of the QCD vacuum
\cite{ACDvac}. From this analysis a picture emerges that, through a remarkable
process of condensation of large chromomagnetic fields, explains the mysterious
phenomenon of colour confinement. In addition to colour confinement, ACD has so
far scored notable successes in the description of the meson ($q\bar{q}$)
\cite{ACDmes} and baryon ($qqq$) spectra \cite{ACDbar}.

The basic picture that ACD paints of the hadronic world is easily outlined
\cite{ACDhad}: the only sector of the theory that must be analysed
non-perturbatively regards the spectrum and wave functions (in terms of a
minimal number of quarks, antiquarks and gluons) of colourless hadronic states.
Such a Primitive World (PW) of hadronic states is (slightly) modified by the
residual quark-gluon and gluon-gluon interaction, which is obtained by
subtraction of the well-defined confining interaction terms. This perturbative
strategy is expected to account for all three-meson and baryon-baryon-meson
couplings, which give finite widths to the unstable members of the PW (which
comprises zero-width hadronic states). Some preliminary satisfactory results
have recently been achieved in this direction \cite{ACDwid}.

In this letter, on the other hand, we shall bring this perturbative strategy to
bear upon the solution in QCD of the fundamental problem of the nature and
structure of high-energy hadronic diffraction, i.e., the Pomeron. In the
following section we outline the calculation leading to the total hadron-hadron
cross section, which in section~\ref{sec:el} is extended to the $\pT$
dependence in the case of elastic scattering.

\section {The Hadron-Hadron Total Cross-Section in ACD}

As already stated, the Pomeron, being an effective exchange with zero quantum
numbers, is suggestive of gluons, which do carry colour however. Thus, in order
not to ``uncover'' the colour of the initially white states, one must
necessarily think in terms of two-gluon exchange (as the leading mechanism), as
suggested in the Low-Nussinov approach \cite{LowNus}.

In ACD the inherent infra-red problems are under control, owing to the fact
that confinement is incorporated into the scheme from the outset and thus, for
example, the gluon is not allowed to propagate over large distances, as in a
standard perturbative QCD. Moreover, from an analysis of the experimental meson
spectrum, which is well described in ACD \cite{ACDmes}, we also have access to
completely determined wave-functions for the scattering hadronic states (at
least in the meson sector).

\begin{figure}[hbt]
\centering
\epsfysize=5cm
\mbox{\epsfbox{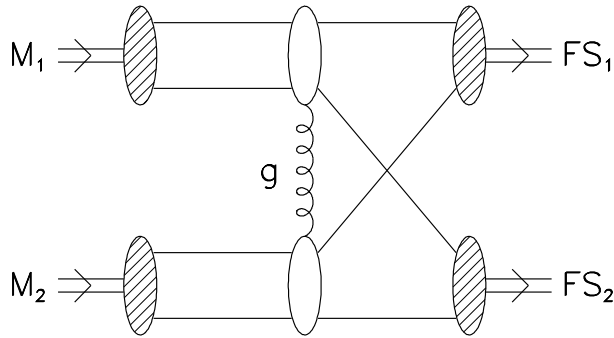}}
\caption{\captionsize Meson scattering into fire-string final states, the
leading contribution to the total cross-section at high energy.}
\label{mmfs}
\end{figure}

The calculation we are thus led to contemplate is represented graphically in
fig.~\ref{mmfs}. In this diagram, M$_{1,2}$ represent the initial-state mesons
and FS$_{1,2}$ the final fire-string states of ACD, which then decay into the
hadronic states detected in the laboratory \cite{Epos}; $g$ is just the ACD
gluon exchange we are considering and the gluon-hadron blobs stand for the two
possible gluon-quark vertices. In terms of hadronic strings, the picture one
then has is an ``entanglement'' of the colliding hadrons due to the
rearrangement of the colour charge, leading to two high-mass fire-strings,
whose decays provide the final hadronic states in the total cross section.

\begin{figure}[hbt]
\centering
\epsfysize=5cm
\mbox{\epsfbox{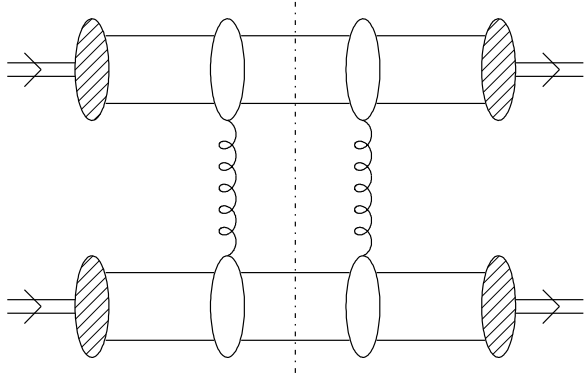}}
\caption{\captionsize The graphical representation for the total meson-meson
cross-section.}
\label{mmtot}
\end{figure}

Since we are not interested in the structure of the final state we may appeal
to the property of completeness and sum over all possibilities and so avail
ourselves of the result \cite{Cea} that one may perform such a calculation
simply by replacing the final states with on-shell quarks. Therefore, squaring
the amplitude, we obtain the cross-section represented in fig.~\ref{mmtot}.

We now have to specify the form of the quark-quark, quark-antiquark potential.
The ACD picture of the hadronic wave-function leads to a system of $q\bar{q}$
confined inside a needle-like domain (in which the colour field is also
confined) of length corresponding to the effective gluon mass of the theory
\cite{ACDvac}. And the potential takes the form

\begin{equation}
 V(r) = \mu^2 r e^{-m_gr},
\end{equation}
where the string tension $\mu$ and the gluon mass $m_g$ are fixed from the
meson-spectrum analysis \cite{ACDmes}:
\begin{equation}
\begin{array}{r@{\;\simeq\;}l}
 \mu & 0.48\gev, \\
 m_g & 0.42\gev.
\end{array}
\end{equation}
Note that there is also an overall sign that is positive (attractive) for
$q\bar{q}$ and negative (repulsive) for $qq$ and $\bar{q}\bar{q}$.

Transforming into momentum space, we obtain an effective potential for such a
gluon exchange of the form
\begin{equation}
 V(Q) = 8\pi\mu^2 \, { (3m_g^2 - Q^2) \over (m_g^2 + Q^2)^3 }.
\end{equation}

For the meson wave-functions we use the following infinite-momentum frame (IMF)
variables:
\begin{eqnarray}
 & & \vec{p_1}=x\vec{p}+\vec\kT \nonumber \\
 \vec p
  \hskip2mm \hbox to0pt{
  \vbox to0pt {\vss
  \hbox to26mm{\hss
  \vbox to25mm{\vss\epsfxsize=24mm
                   \epsfbox{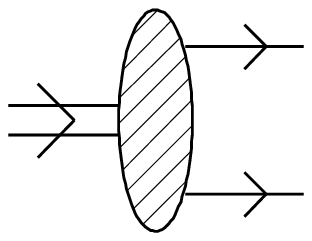}\vskip5mm}
               \hss}
               \vss}
               \hss}
  \hskip20mm & & \hskip38mm
 =\phi^{\vec{p}}(x,\kT), \\
 & & \vec{p_2}=(1-x)\vec{p}-\vec\kT \nonumber
\end{eqnarray}
where one has the usual restrictions $0\le x\le1$, $|\kT|\ll p$. In this frame,
with the quarks approximately on-shell, it is easy to show that energy
conservation demands that the longitudinal momentum fraction carried by the
gluon be ${\approx}O(\kT^2/p^2)$ (i.e., vanishingly small in the IMF) and so
may be neglected. With this approximation the calculation simplifies
considerably, as all longitudinal dynamics drops out.

Thus, we obtain for the total cross-section
\begin{equation}
 \begin{array}{rl}
 \sigma_{\pi\pi}^{\mathrm{tot}}
 =&
 {\fcol \over (2\pi)^6} \;
 \int d^2\kT \, d^2\kT' \, \half d^2\QT \; \left|V(\QT)\right|^2 \nonumber \\
 &\times\!
  \left|\phi_\Trans(\kT {-}\half\QT)-\phi_\Trans(\kT {+}\half\QT)\right|^2 \,
  \left|\phi_\Trans(\kT'{-}\half\QT)-\phi_\Trans(\kT'{+}\half\QT)\right|^2 \! ,
 \end{array}
\end{equation}
where $\fcol=2/9$ is the colour factor from tracing the vertex colour matrices.

We parametrize the transverse part of the wave-functions with
\begin{equation}
 \phi_\Trans(\kT) = N \exp \left( -{\kT^2 \over 2\kappa^2} \right),
\end{equation}
where the normalisation fixes $N$ and the transverse-momentum cut-off parameter
$\kappa$, as obtained from the meson spectrum analysis \cite{ACDmes}, is
$\kappa\simeq0.15\gev$. So the $\kT$ and $\kT'$ integrals may be performed
analytically and we arrive at
\begin{equation}
 \sigma_{\pi\pi}^{\mathrm{tot}} = {\fcol \over 2(2\pi)^2} \; \int d^2\QT
 \; \left| V(\QT^2) \right|^2 \; 4 \left( 1 - e^{-\QT^2/4\kappa^2} \right)^2,
\end{equation}
This final integral may be performed numerically to give an answer of the form
\begin{equation}
 \sigma_{\pi\pi}^{\mathrm{tot}} = 32\pi \, \fcol \, I(\xi) \,
 \left({\mu\over m_g}\right)^4 \, {1\over m_g^2},
\end{equation}
with
\begin{equation}
 I(\xi) = \int_0^\infty dy \;
 (1-e^{-\xi y})^2 \, \left[(3-y)\over(1+y)^3\right]^2,
\end{equation}
where $\xi=m_g^2/4\kappa^2$ and $y=\QT^2/m_g^2$.
Thus, we finally obtain
\begin{equation}
 \begin{array}{rcl}
 \sigma_{\pi\pi}^{\mathrm{tot}}
 &\simeq&
 0.09\pi \, \fcol \, \left({\mu\over\kappa}\right)^4 \, {1\over m_g^2} \\
 &\simeq& 15\nb.
 \end{array}
\end{equation}
This figure is to be compared with the experimentally accessible
\begin{equation}
 \sigma_{pp}^{\mathrm{tot}}    \simeq 39\nb \quad \mathrm{and} \quad
 \sigma_{\pi p}^{\mathrm{tot}} \simeq 24\nb.
\end{equation}
Note that if one assumes the factorization implicit in our calculation then
these numbers are perfectly compatible, since one then expects
\begin{equation}
 \sigma_{pp}^{\mathrm{tot}} \; \cdot \; \sigma_{\pi\pi}^{\mathrm{tot}}
 = \left( \sigma_{\pi p}^{\mathrm{tot}} \right)^2.
\end{equation}

So far, we have considered the final-state fire-strings as having an
energy-independent transverse radius. However, owing to the logarithmic growth
in multiplicity \cite{Epos} and the consequent logarithmic growth of the
transverse momentum ``kicks'' in the decay process, there will be a logarithmic
dependence on energy of the transverse size of the fire-string states. Such a
growth spoils the na{\"\i}ve approximation \cite{Cea} necessary to arrive at
the sum of quark states and a residual $\log{s}$ dependence is introduced into
the amplitude. Unfortunately we are not at present able to perform the
calculations necessary to obtain a precise parametrisation of this growth,
although we note that the implied $\log^2s$ growth of the total cross section
represents well the available experimental data.

\section {The Elastic Scattering Cross-Section}
\label{sec:el}

We can take this calculation a step further: by using the optical theorem, the
diagram of fig.~\ref{mmtot} is just the imaginary part of the forward elastic
$\pi\pi$ cross-section. If the outgoing pion pair are now assigned a different
momentum, $p'$ say, then writing $p'=xp+\pT$ we see that for $\pT$ small
$x{\approx}1$ and the only alteration is that the two exchanged gluons now have
different transverse momenta. The calculation is then a straightforward
modification of the previous and one easily arrives at
\begin{equation}
 \begin{array}{rl}
 \mathrm{Im} \, A_{\pi\pi}^{\mathrm{el}}(\pT) = {\fcol \over 2(2\pi)^2}
 \int d^2\QT & V(\QT{-}\half\kT) \, V^\dagger(\QT{+}\half\kT) \\
             & \times \;
  4 \left[ e^{-({1\over2}\kT)^2/4\kappa^2} - e^{-\QT^2/4\kappa^2} \right]^2,
 \label{sigel}
 \end{array}
\end{equation}

\begin{figure}[hbt]
\centering
\epsfysize=10cm
\mbox{\epsfbox{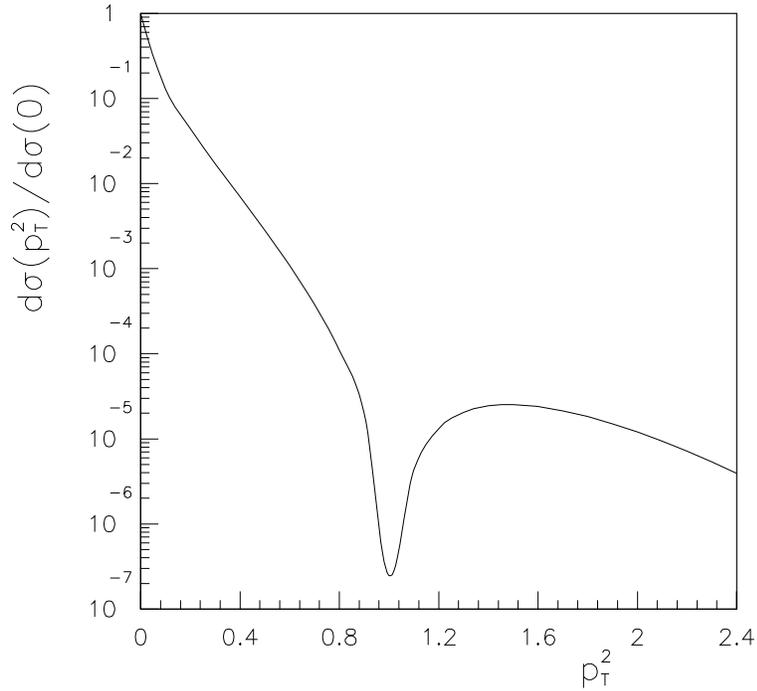}}
\caption{\captionsize Our results for the $\pi\pi$ differential elastic
cross-section as a function of $\pT^2$.}
\label{elpipi}
\end{figure}

In fig.~\ref{elpipi} we display the $\pi\pi$ differential elastic cross-section
obtained in our calculation as a function of $\pT^2$. Several features are
noteworthy. First and foremost, the striking dip at around $1\gev$, which is
well-established in the case of high-energy $pp$ scattering at a similar value
of $\pT$. Less obvious, but equally established experimentally, is the break in
the slope parameter in the region of $\pT^2\sim0.2\gev^2$, above this value and
below the dip a parameterization of the form $\exp(-b\pT^2)$ gives
$b\sim10\gev^{-2}$. Asymptotically, for large $\pT$, the behaviour takes on the
form of a power law $\sim\pT^{-8}$.

The transparency of the calculations, performed in the infinite-momentum frame,
make it rather easy to elucidate the origin of these effects. The finite range
of the potential, due to confinement, implies the existence of zeroes in the
scattering amplitude starting at a value of $\pT$ of the order of
$\pi/r_{\mathrm{conf}}$, where the confinement radius
$r_{\mathrm{conf}}\sim(0.4\gev)^{-1}$ as discussed above. The higher zeroes
are, of course, to a large extent smeared out in the integral, a full treatment
including an Eikonal re-summation may eventually also lead to some smearing of
the primary zero. The form of the final integrand in $\QT$, see
eq.~\ref{sigel}, then explains the lower $\pT$ behaviour: for very low $\pT$
the behaviour is dominated by the first exponential in eq.~\ref{sigel} and thus
by the width of the $\pT$ distribution (obtained from the meson spectrum
analysis), this is then overtaken by the second term, which leads to a partial
cancellation that is, however, never clearly evident due to the very steep
fall-off of the amplitude in this region and only the slight break in slope
parameter remains. The large-$\pT$ behaviour is dictated entirely by the form
of the potential used and is seen to be approximately of the form $\pT^{-8}$.

% ----------------------------- END OF MAIN TEXT ------------------------------

\end{document}